# Evaluation of optical constants in oxide thin films using machine learning

*Kyosuke Saeki[1] and Takayuki Makino[2,1,*]*

[1] *Department of Electric and Electronics Engineering, University of Fukui, Fukui 910-8507, Japan*

[2] *Research Center for Development of Far-Infrared Region, University of Fukui, Fukui 910-8507, Japan*

E-mail: tmakino@u-fukui.ac.jp

Abstract This paper describes an inverse analysis method using neural networks on optical spectroscopy, and its application to the quantitative optical constant evaluation. The present method consists of three subprocesses. First, measurable UV-visible spectroscopic quantities were calculated as functions of the optical constants of the solid based on the Tomlin equations [J. Phys. D 1 1667 (1968)] by carefully eliminating the unpractical combinations of optical constants. Second, the back-propagation neural network is trained using the calculated relationships between the measurable quantities and the optical constants. Finally, the trained network is utilized to determine the optical constants from measured responses. The conventional (Newton-Raphson) method tends to require the judgement of a well-experienced analyst, while machine learning shows automatically human-free performance in data conversion.

# 1. Introduction

In recent years, the importance of elucidating basic physical properties has been increasingly recognized in the development of organic and inorganic semiconductors. Efforts are being made to gain a deeper understanding of the electronic structure. The importance of optical evaluation as a characterization tool for this is well known. One of the representative techniques for such an optical characterization is UV-visible spectroscopy (UVS). A transformation of the measured quantities into optical constants (complex refractive index) is often performed to facilitate comparison with the electronic structure properties. However, it is known that the conversion of these physical quantities may become an inverse problem depending on a semiconductor structure to be measured. Even worse, there is a multi-solution problem. By setting $R$ is the reflectance and $T$ transmissivity, let us consider light of wavelength $\lambda$ in air or vacuum incident normally on a film of thickness $d$ and complex refractive index $\eta - i\kappa$ which is supported on a non-absorbing thick substrate of refractive index $n_s$.

The expressions of $R$ and $T$ are too complicated to be shown even in this case. Following Heavens [1], quantities of $R$, $T$ are given by,

$$R = \frac{a_1 \exp(2d_2) + b_1 \cos(2d_1) + c_1 \sin(2d_2) + f_1 \exp(-2d_2)}{a_2 \exp(2d_2) + b_2 \cos(2d_1) + c_2 \sin(2d_2) + f_2 \exp(-2d_2)} \quad (1)$$

$$T = \frac{32 n_s^2 (\eta^2 + \kappa^2)}{a_2 \exp(2d_2) + b_2 \cos(2d_1) + c_2 \sin(2d_2) + f_2 \exp(-2d_2)} \quad (2)$$

where

$$a_1 = ((\eta - 1)^2 + \kappa^2)\big((n_s^2 + 1)(\eta^2 + \kappa^2 + n_s^2) + 4\eta n_s^2\big) \quad (3)$$
$$a_2 = ((\eta + 1)^2 + \kappa^2)\big((n_s^2 + 1)(\eta^2 + \kappa^2 + n_s^2) + 4\eta n_s^2\big) \quad (4)$$
$$b_1 = -2\big((n_s^2 + 1)(\eta^2 + \kappa^2 - 1)(\eta^2 + \kappa^2 - n_s^2) + 8\eta n_s^2\big) \quad (5)$$
$$b_2 = -2\big((n_s^2 + 1)(\eta^2 + \kappa^2 - 1)(\eta^2 + \kappa^2 - n_s^2) - 8\eta n_s^2\big) \quad (6)$$
$$c_1 = 4\kappa\big(-(n_s^2 + 1)(\eta^2 + \kappa^2 - n_s^2) + 2n_s^2(\eta^2 + \kappa^2 - 1)\big) \quad (7)$$
$$c_2 = 4\kappa\big((n_s^2 + 1)(\eta^2 + \kappa^2 - n_s^2) + 2n_s^2(\eta^2 + \kappa^2 - 1)\big) \quad (8)$$
$$f_1 = ((\eta + 1)^2 + \kappa^2)\big((n_s^2 + 1)(\eta^2 + \kappa^2 + n_s^2) - 4\eta n_s^2\big) \quad (9)$$
$$f_2 = ((\eta - 1)^2 + \kappa^2)\big((n_s^2 + 1)(\eta^2 + \kappa^2 + n_s^2) - 4\eta n_s^2\big) \quad (10)$$

with

$d_1 = 2\pi\eta d/\lambda$, and $d_2 = 2\pi\kappa d/\lambda$.

The measurable quantities $R$ and $T$ are usually represented as functions of complex refractive indices ($\eta$ and $\kappa$). This is regarded as a direct problem. Different from the case of a substrate-free standalone wafer, the optical constants ($\eta$ and $\kappa$) cannot be represented by these quantities ($R$ and $T$). Because of the intimacy of the optical constants with the electronic structure of the compound, conversion of the measured quantities into these is often performed. On the other hand, this is classified into an inverse analysis.

Traditionally, this type of conversion has been solved numerically with *e.g.* Newton's method. Even worse, we suffer from the multi-solution problem as indicated by Fig. 1. When we try to find the root from the $\eta - \kappa$ mapping, we usually look for intersections between $|R - R_m| \simeq 0$ and $|T - T_m| \simeq 0$, where $R_m$ and $T_m$ are measured reflectance and transmittance, respectively. As is indicated in this figure, sometimes there are multiple intersections in the mapping. If an analyst is well-experienced, he or she can choose the appropriate solution from these based on knowledge on impractical combinations of $\eta$ and extincition coefficient $\kappa$, for example, excessively small $\eta$ despite sufficiently high extinction coefficient, high $\kappa$ or high $\eta$ in the transparent wavelength regions. In the case making a computer perform this task, adequateness tends to severely depend on the selection of the initial values. A well-experience analyst could prepare an appropriate initial value set for the correct choice, which may hinder the development of automated analytical methods. One can consider about an implementation of such a restricted optimization problem in the traditional framework using *e.g.* penalty functions. Nevertheless, the code would become very difficult and complicated to implement.

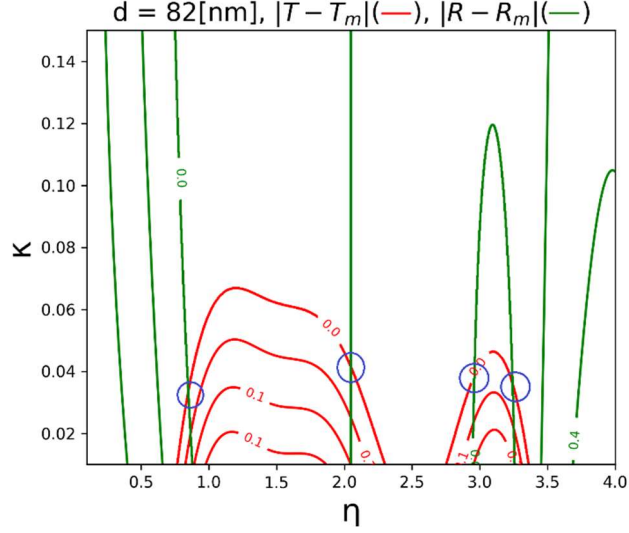

**Fig. 1** Contour-plot representations of $|T - T_m|$(red), $|R - R_m|$(green) as functions of complex refractive indices, $\eta$ and $\kappa$ with subsrate's refractive index of 1.768 and the film thickness of 82 nm. The material is ε-Ga$_2$O$_3$.

Traditionally, dielectric functions have been obtained by iterating the Newton-Raphson method to minimize an error function consisting of calculated and experimental values, without using a model dielectric function (MDF). It is now widely adopted that spectroscopic ellipsometric analysis relies on the MDF, which is often found to be easier to perform. [2]

A development of the MDF-free approach is that by the Likhachev group. [3–5] There, the problem is approached with a B-spline representation of the dielectric function.

Here, we describe about candidates to avoid the abovementioned multi-solution problem.

One possible way is the thickness-dependent measurement by preparing several samples. Denton and coworkers [6] demonstrated that the wavelengths with the multi-solution depend on the film thickness. Preparation of a sufficiently large number of the samples with different thicknesses may allow us to avoid the multiple solutions. This is not a good way to work it around from the cost-performance point of view. Equivalently, one can change the angle of incidence instead of changing the thickness. However, changing the incident angle tends to induce unwanted polarization rotation and ellipticity effects. [7–9]

Extensive studies on the inversion problems have been conducted also in other disciplines such as a computer mechanics. [10–14] Attempt on the structural identification of a beam is made with measured eigen-frequencies and eigen-modes. This conversion utilizes machine-learning-related neural-network. [15–21] In contrast to recent MDF-based transformations to dielectric functions, neural network-based transformations have the advantage of being MDF-free and automatic. Even better, during the construction of the training data set, one can include the probable combination of the complex refractive indices based on *e.g.* model dielectric functions.

In order to solve the above-mentioned current problems, this paper aims to establish an implementation method for automatic analysis of UVS and examines the effectiveness of supervised machine learning methods that have recently developed rapidly. [22–25]

## 2 Experimental procedures

In this work, UV-Vis spectrophotometer was used to measure the transmission $T$ and reflectance $R$. A light source of deuterium and halogen lamps fires an incident beam in the wavelength range of 190 to 900 nm. Monochromatized light then passes through or reflects on a sample. After that, a photomultiplier tube is used to detect the transmitted or reflected light as a function of wavelength $\lambda$ or photon energy. Details can be found elsewhere. [1]

## 3 Calculation procedures

3.1 Structure of neural networks

A neural network consists of several layers: an input, an intermediate, and an output layer. This unit models a neuron, which receives multiple input values and outputs multiple values. Each input is multiplied by a weight, and a constant called the bias is added to the sum of the weights. The product of the inputs and weights plus the bias is processed by the activation function. Here, we used a sigmoid function for the middle layer activation function and a constant function for the output layer activation function. If the number of units in the input, intermediate, and output layers are $l$, $m$, and $n$, respectively, the output of $j$ units in the intermediate layer is given by

$$u_j = f\left(\sum_{i=1}^{l} x_i w_{ij} + b_j\right) \tag{11}$$

$$y_j = \frac{1}{\{1 + \exp(-u_j)\}} \tag{12}$$

where $x_i$ is the input from input layer $i$th unit to intermediate layer $j$th unit, $w_{ji}$ is the weight between the input and intermediate layer units, and $b_j$ is the bias of intermediate layer $j$th unit.

There are no weights between units in the same layer, but weights exist between two adjacent unit layers. Figure 2 shows an ordinary three-layer neural network.[26] it has been proven that a three-layer neural network can represent any nonlinear mapping. Therefore, a three-layer neural network was employed here.

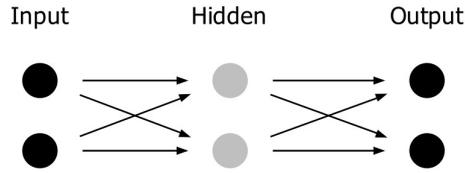

**Fig. 2** Schematic illustration of neural network used in the current work.

The basic idea of learning a neural network is as follows. First, we define an error function $E$ as follows.

$$E = \frac{1}{2} \sum_k (y_k - t_k)^2 \tag{13}$$

where $E$ is an error, $y_k$ is an output value of the output layer, and $t_k$ is a correct answer value.

In the learning process, the weights $w_{ji}$ and bias $b_j$ are iteratively updated based on the stochastic gradient descent method [27] to minimize the above error. This learning allows the neural network to output values similar to the correct answer value. This learning algorithm is called backpropagation.[28] the stochastic gradient descent method is characterized by the fact that samples are randomly selected from the training data for each update, making it difficult to be trapped in a local solution. The amount of update is determined by multiplying the gradient of the error function by the learning coefficient. The update formula for the stochastic gradient descent method is defined by the following equations.

$$w \leftarrow w - \eta \frac{\partial E}{\partial w} \tag{14}$$

$$b \leftarrow b - \eta \frac{\partial E}{\partial b} \tag{15}$$

Here, $\eta$ is the learning coefficient.

The number of units in the input and output layers is automatically determined by the problem. The output layer has two units whose outputs are the real and imaginary parts ($\varepsilon_1$, $\varepsilon_2$) of the dielectric function. The relationship between complex refractive indices and dielectric functions is as follows:

$$\eta = \frac{1}{\sqrt{2}}\left(\varepsilon_1 + (\varepsilon_1^2 + \varepsilon_2^2)^{\frac{1}{2}}\right)^{\frac{1}{2}} \tag{16}$$

$$\kappa = \frac{1}{\sqrt{2}}\left(-\varepsilon_1 + (\varepsilon_1^2 + \varepsilon_2^2)^{\frac{1}{2}}\right)^{\frac{1}{2}} \tag{17}$$

However, the number of units in the hidden layer varies from problem to problem. Since the backpropagation is based on the stochastic gradient descent, the weights and bias values have a significant impact on the learning and estimation. In addition, the error value usually monotonically increases as the number of training iterations of the training data decreases. However, during the estimation phase, the error phenomenon sometimes stops and increases. This is called overlearning. [29] Therefore, in this study, two units were employed in the hidden layer after testing several calculations.

### 3.2 Tomlin equations

Tomlin [30] found that changing the quantities of left-hand sides of Eqs. (1) and (2) to $(1 \pm R)/T$ significantly simplifies the complexity of the equations. Thus, we adopt this version as inputs of training data, not $T$ nor $R$, directly. here, which read: [30]

$$\frac{1+R}{T} = \frac{(1+\eta^2+\kappa^2)\big((\eta^2+n_s^2+\kappa^2)\cosh(2\alpha_1) + 2\eta n_s \sinh(2\alpha_1)\big)}{4n_s(\eta^2+\kappa^2)}$$
$$+ \frac{(1-\eta^2-\kappa^2)\big((\eta^2-n_s^2+\kappa^2)\cos(2\gamma_1) - 2(n_s\kappa)\sin(2\gamma_1)\big)}{4n_s(\eta^2+\kappa^2)} \tag{18}$$

$$\frac{1-R}{T} = \frac{\eta\big((\eta^2+n_s^2+\kappa^2)\sinh(2\alpha_1) + 2(\eta n_s)\cosh(2\alpha_1)\big)}{2n_s(\eta^2+\kappa^2)}$$
$$+ \frac{\kappa\big((\eta^2-n_s^2+\kappa^2)\sin(2\gamma_1) + 2(n_s\kappa)\cos(2\gamma_1)\big)}{2n_s(\eta^2+\kappa^2)} \tag{19}$$

### 3.3 Model dielectric function

One option would be to calculate and obtain the measured quantities for all possible combinations of $\eta$ and $\kappa$ as far as is practically possible ($0 \leq \eta \leq 8,\ 0 \leq \kappa \leq 0.8$), but because of the multi-solution problem described above, a different method is employed here, as described below. The adopted method relies on the model dielectric functions, which will be explained later in detail. Transitions involving these energy bands are responsible for all the features in the dielectric functions. In a model that we have employed, one approximates the dielectric function as a sum of several components, each of which is an explicit function of energy($\hbar\omega$). The dielectric function represents a contribution from the neighborhood of a critical point in the joint density of states. [31] We use the following $M_0$-type dielectric functions for the $E_0$ (the band gap energy) transition. This transition occurs at a photon energy of 4.89 eV at 150 K for ε-Ga$_2$O$_3$. Assuming the bands are parabolic, we obtain the contribution of this gap to the dielectric function [32,33]:

$$\varepsilon(\hbar\omega) = \frac{A_0}{E_0^{1.5}} \times \left(\chi_0^{-2}(2 - (1+\chi_0)^{0.5} - (1-\chi_0)^{0.5})\right) \quad (20)$$

with

$$\chi_0 = \frac{\hbar\omega + i\Gamma}{E_0} \quad (21)$$

values of $A_0$ and $\Gamma_0$ in Eqs. (20) and (21) correspond to the strength and broadening parameters in this optical transition.

We next describe the $E_1$ feature. We labeled a structure found in the optical spectra in the region higher than $E_0$ as $E_1$ (~ 6.8 eV). The $E_1$ peak is difficult to analyze as it does not correspond to a single, well-defined critical point. Thus, we have characterized the $E_1$ structure as a damped harmonic oscillator [32,33]:

$$\varepsilon(\hbar\omega) = \frac{A_1}{1 - \chi_1^2 - i\Gamma_1\chi_1} \quad (22)$$

with $\chi_1 = E/E_1$, where $A_1$ is a *dimensionless* strength parameter. The quantity $\Gamma_1$ is a *dimensionless* broadening parameter.

We then describe the total dielectric function. The whole dielectric function was found by summing the expressions given previously. We created our training dataset by changing the parameters in the dielectric function such as $A_0$ and $A_1$ from minimum to maximum values shown in Tables I and II. If we wish to obtain, for example, the imaginary parts $\varepsilon_1$, we can take the imaginary part $\varepsilon_2$ of Eqs. (20) and (22). A constant, $\varepsilon_{1\infty}$ was added to the real part of the dielectric constant $\varepsilon_1$ to account for the vacuum plus contributions from higher-lying energy gaps. [32–34] Figure 3 shows the individual contributions to the real $\varepsilon_1$ and imaginary $\varepsilon_2$ parts of the complex dielectric functions, respectively, of the two transitions.

**Table I.** Material parameters used in the calculation of optical constants for ε-Ga$_2$O$_3$.

|         | $\varepsilon_{1\infty}$ (eV) | $E_0$ (eV) | $\Gamma_0$ (eV) | $A_0$ (eV$^{3/2}$) | $E_1$ (eV) | $\Gamma_1$ (eV) | $A_1$ (eV$^{3/2}$) |
|---------|------|------|------|----|-----|------|-----|
| minimum | 1.2  | 4.86 | 0.02 | 30 | 6.7 | 0.05 | 0.4 |
| maximum |      | 4.89 | 0.04 | 50 | 6.8 | 0.07 | 0.6 |
| increment |    | 0.01 | 0.01 | 10 | 0.025 | 0.01 | 0.1 |

**Table II.** Material parameters used in the calculation of optical constants for Yb$_2$O$_3$.

|         | $\varepsilon_{1\infty}$ (eV) | $E_0$ (eV) | $\Gamma_0$ (eV) | $A_0$ (eV$^{3/2}$) | $E_1$ (eV) | $\Gamma_1$ (eV) | $A_1$ (eV$^{3/2}$) |
|---------|------|------|------|----|-----|------|-----|
| minimum | 1.2  | 4.62 | 0.03 | 30 | 6.7 | 0.07 | 0.4 |
| maximum |      | 4.66 | 0.05 | 50 | 6.8 | 0.09 | 0.6 |
| increment |    | 0.01 | 0.01 | 10 | 0.025 | 0.01 | 0.1 |

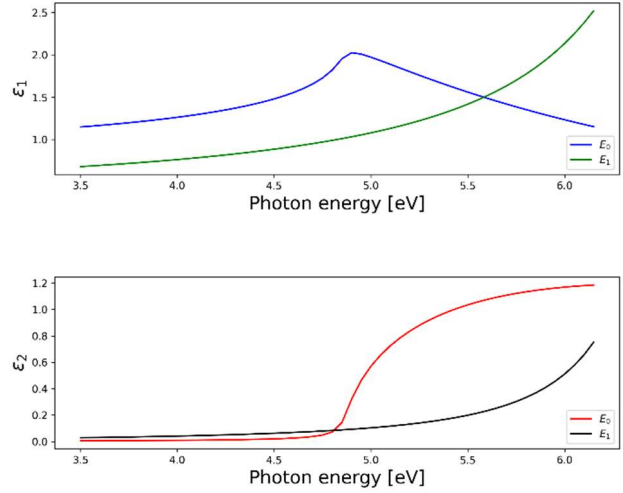

**Fig. 3** Model dielectric functions used to create the training data sets for ε-Ga$_2$O$_3$, and $E_0 = 4.88$ eV, $E_1 = 6.8$ eV.

3.4 Basic principles of current Inverse problem analysis

The evaluation of optical properties by UVS involves the conversion of measured quantities to dielectric functions. The following inverse analysis procedure using a neural network is considered in this study.

1. the measured UVS quantity is calculated as a function of the dielectric function based on the Tomlin equation and the MDF calculation. These values are used as training data.

2. train a neural network using the training data obtained in 1.

Here, $(1 \pm R)/T$ are used as inputs, and the real $\varepsilon_1$ and imaginary $\varepsilon_2$ parts of the dielectric function are used as outputs.

3. using the trained neural network, we calculated the dielectric function from the measured quantities ($R$ and $T$).

## 4. Result and discussion

We here show examples of the converted results on ε-Ga$_2$O$_3$ and Yb$_2$O$_3$ respectively.

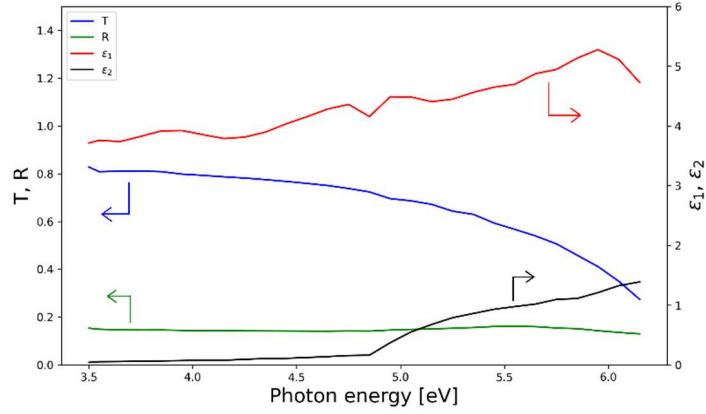

**Fig. 4** Transmissivity $T$, reflectance $R$, converted complex dielectric functions ($\varepsilon_1$ and $\varepsilon_2$) of ε-Ga$_2$O$_3$ at 150 K.

Figure 4 shows $T$ and $R$ spectra obtained by our measurements for ε-Ga$_2$O$_3$ film [35] grown on sapphire substrate at 150 K. Also, we show in Fig. 4 the converted complex dielectric functions $\tilde{\varepsilon}$, through the abovementioned calculation using the neural network.

We now discuss the validity of the transformed results. We notice that the lineshape of $\varepsilon_2$ spectrum is characterized by the clear onset at $E_g$, which is inhered to the typical behavior of a widegap semiconductor. Moreover, we confirm on one hand, an optical anomaly at $E_g$ in $\varepsilon_1$ spectrum. Our present results are similar to those reported previously for the identical. [35] It is known that the accuracy of $\varepsilon_{1\infty}$ has a great influence on the real part of the dielectric function. Adachi and Taguchi [34] indeed mentioned about the difficulty of evaluating $\varepsilon_{1\infty}$ experimentally from the spectral lineshape analysis on ZnSe single crystal. [34] On the contrary, we here expect the DFT calculation [36,37] may have a sufficient accuracy in results. We think that the coincidence on the comparable scale with the conversion results might be an evidence that our adopted method is reasonable. We adopted classical NR method to reduce the $\tilde{\varepsilon}$ spectra. We wish to compare our result with the first principles DFT calculation. Figure 5 shows comparison of the complex dielectric functions: experiments vs. GW-level calculations in ε-Ga$_2$O$_3$. DFT $\varepsilon_1$ and $\varepsilon_2$ shifted by 2.0eV. In general, the spectroscopic line shape of the dielectric function is determined by height, band gap energy, and broadening. First-principles electronic structure calculations or density functional theory are well-known for underestimation of the band gap energy. Therefore, we shifted this. Here we decided the energy shift based on the dip position in the transmittance spectrum. These results are in a reasonably good agreement each other. It should be noted that one of the line shape factors, is, in other words. In first-principles electronic structure calculations, one introduces broadening as a phenomenological parameter. The value of broadening used in the theoretical calculation is as large as 0.2eV, which is significantly larger than the realistic broadening of *ca.* 0.05eV. this might be a reason of discrepancy from the lineshape point of view.

Broadening parameter deduced from the model analysis was *ca*. 50 meV. The phenomenological broadening in the theoretical calculations is significantly larger than the actual one, so the steepness at the rise of the imaginary dielectric function at the band gap is loose in the theoretical results.

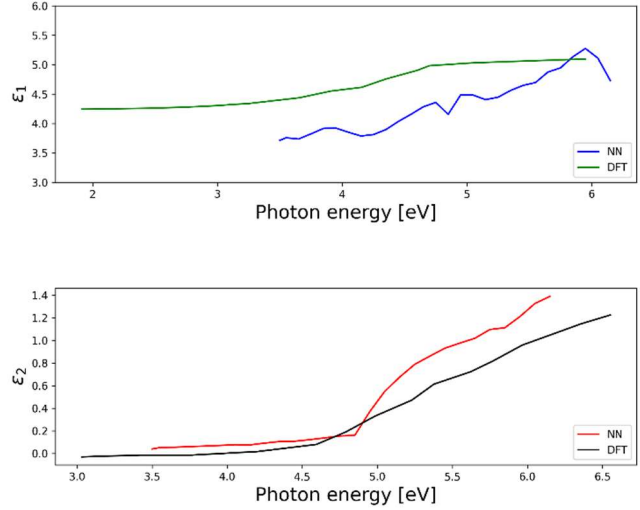

**Fig. 5** Comparison of the complex dielectric functions: experiments (NN) vs. GW-level calculations (DFT) in ε-$Ga_2O_3$. DFT results were shifted by 2.0 eV.

We make a discussion here regarding universality. Figure 6 shows $T$, $R$, converted complex dielectric functions $\tilde{\varepsilon}$ of $Yb_2O_3$ [38,39] at room temperature. We again wish to emphasize the reproducible behavior as characterized an onset at $E_g$ (band gap energy) in $\varepsilon_2$ spectrum and optical anomaly at $E_g$ in $\varepsilon_1$ spectrum typical behavior as semiconductor spectrum similar to that summarized in the our previous results. [39]

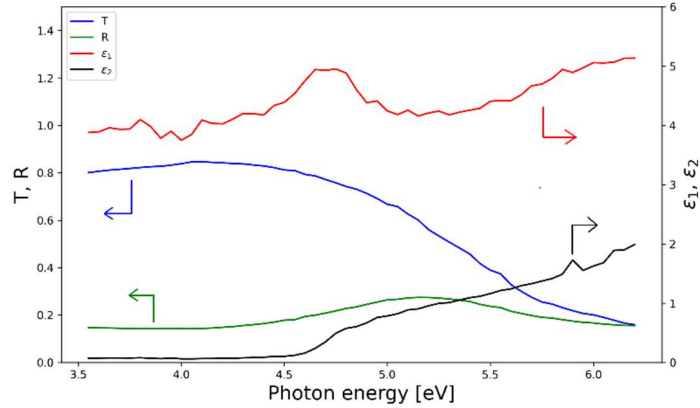

**Fig. 6** Transmissivity $T$, reflectance $R$, converted complex dielectric functions ($\varepsilon_1$ and $\varepsilon_2$) of $Yb_2O_3$ at room temperature.

## 5. Conclusion

In this paper, we present an inverse analysis method using neural networks and the evaluation of quantitative optical constants. An attempt to automatically convert measurable UVS quantities into optical constants, based on machine learning methods, allowed us to validate the effectiveness of the machine learning method. We also checked versatility of this proposed method by assessing two materials, *i.e.*, ε-$Ga_2O_3$, $Yb_2O_3$ respectively.

## Acknowledgements

We are indebted to T. Asai and T. Takeuchi for permission to use some of their experimental results. The authors are pleased to acknowledge that the specimens used in this study was provided by D. Oka and T. Fukumura. A partial financial support of this work from JSPS KAKENHI Grant Number 19K05303 was also acknowledged. A part of this work was conducted with the use of the UV–vis spectrometer. We also acknowledge scientific supports by N. Kuzuu who own this equipment.